\documentclass[prl,floats,twocolumn,twoside,floatfix,amsmath,amssymb,]{revtex4}

\usepackage{epsfig}

\newcommand{\taurel}{\tau_{\scriptscriptstyle \rm rel}}

\begin{document}

\title{Asymmetric exclusion processes with constrained dynamics}
\author{ 
M. Sellitto\thanks{E-mail: \email{sellitto@ictp.it}} 
}
\affiliation{The Abdus Salam International Centre for
Theoretical Physics, \\ Strada Costiera 11, 34100 Trieste, Italy}

\begin{abstract}
  Asymmetric exclusion processes with locally reversible kinetic
  constraints are introduced to investigate the effect of
  non-conservative driving forces in athermal systems.  At high
  density they generally exhibit rheological-like behavior, negative
  differential resistance, two-step structural relaxation, dynamical
  heterogeneity and, possibly, a jamming transition driven by the
  external field.
\end{abstract}


\maketitle

There is a growing appreciation that glassy relaxation can be ascribed
to purely dynamic restriction on the particle motion with static
correlations and related thermodynamic factors playing little or no
role~\cite{ToBiFi,Paolo,Merolle}.  Kinetically constrained models
(KCMs) provide a simple and elegant way to rationalize this idea and
have spurred the challenge to reproduce much of the observed glassy
behavior~\cite{KCM_reviews}, including detailed predictions that first
originated in apparently unrelated and complementary mean-field
approaches~\cite{Se_review,SeBiTo}.  While investigations of KCMs have
been mostly focused on equilibrium and aging dynamics, there are
relatively few studies dealing with the effects of nonconservative
forces.  The issue is of special interest in rheology, where the
apparent viscosity of structured liquids is found to depend on the
applied stirring force~\cite{Larson}.  Shear-thinning refers to the
commonly observed situations in which the viscosity decreases at
increasing forces, and is
well described by mode-coupling theory~\cite{BeBaKu}.  The opposite,
shear-thickening, behavior is less common and much more difficult to
predict.  In some cases the viscosity increase can be so dramatic that
the macroscopic motion may even stop, the liquid
jams~\cite{Larson,bibette}.  It has been suggested that
shear-thickening and jamming~\cite{note_jamming} are related to an
underlying glass transition~\cite{Holmes}.  The idea has been
interpreted microscopically in terms of an entropy-driven inverse
freezing~\cite{SeKu} and recent experiments have come to support this
interpretation~\cite{Bonn,Roberta}.
The latter approach, however, requires some special thermodynamic and
structural features which are absent in athermal shear-thickening
systems (e.g. concentrated suspensions of hard spheres).  Accordingly,
any attempt at unifying different types of dynamical arrest in systems
dominated by steric hindrance and packing effects should explain the
simultaneous emergence of thinning and thickening behavior with their
underlying glassy dynamics.

In this Letter I will show that rheological-like behavior occurs in
microscopic models of finite dimensional particle systems interacting
only through non-Hamiltonian forces and purely kinetic constraints.
Evidence is provided by considering a variant of the asymmetric simple
exclusion process (ASEP)~\cite{ASEP_reviews}, in which particle motion
obeys an additional constraint motivated by lattice glass
models~\cite{KoAn}.  Similarly to the ASEP the transition
probabilities satisfy local detailed balance although, due to the
periodic boundary condition, the driving field cannot be derived from
a Hamiltonian.
The applied field induces a nonequilibrium steady state (NESS) in the
system.
At small field the transport is ohmic and thinning behavior is
observed.  At high density, due to the imposed kinetic restrictions on
the particle motion, the ohmic regime shrinks and the current
decreases at increasing field, a negative differential resistance
effect.  The system can thus be driven, at constant density, into a
regime close to jamming in which the dynamics is slow but stationary.
In this regime, shear-thickening and other salient features of glassy
behavior like two-step relaxation and dynamical heterogeneities are
observed.

\paragraph{The model.}

ASEP describes a 1D lattice model of hard-core particles hopping
randomly to a vacant nearest-neighbour site with rates $p_{\pm}$
depending on the direction of the particle move. Current-carrying
NESSs are maintained either by fixing periodic boundary conditions, or
by connecting the open boundaries to two particle-reservoirs at
different densities.  In the latter form the ASEP was originally
introduced to model the biopolymerization kinetics on nucleic acid
templates~\cite{acid}, and has come to play a paradigmatic role in
recent developments of nonequilibrium statistical
mechanics~\cite{ASEP_reviews,Derrida_review}.  ASEP-type models have
been also studied in relation to vehicular traffic, molecular motors
and intra/extra-cellular transport~\cite{Andreas}.

In this work, a variant of the 2D ASEP with discrete-time evolution is
introduced in which particle hopping occurs with probability $p={\rm
  min} \{1,\, \exp{(\vec{E} \cdot \vec {dr}) } \} $, where $\vec{E}$ is
the applied field and $\vec{dr}$ is the displacement unit vector,
provided that an additional condition is met: the particle is
constrained to have a number of neighboring particles below a certain
threshold both before and after the move~\cite{KoAn}.  The constraint
is naturally inspired by the cage effect in viscous liquids and leads
to glassy relaxation at high density.
In zero field,  one recovers the boundary-driven constrained
diffusion model of Ref.~\cite{Se_ratchet}. For periodic boundary
conditions, the kinetic nature of the constraint guarantees that the
equilibrium measure is trivial.  The fully irreversible case is
obtained in the opposite limit of infinite drive, corresponding
to the kinetically constrained version of the totally asymmetric
simple exclusion process. 

For simplicity we consider square-lattice systems (of size $L^{2}$)
with periodic boundary conditions in which the driving force is
applied along a lattice axis.
The kinetic constraint is chosen to depend isotropically on the
nearest neighbors with a threshold set to three.  This constraint
ensures that diffusion coefficient at equilibrium is finite at any
nonzero vacancy~\cite{ToBiFi}.  Thus, no ``true'' glass transition is
present in zero field although the dynamics is slow at high density
and characterized by the cooperative motion of spatially extended
mobility defects, whose size grows with the density~\cite{ToBiFi}.
For finite field, every allowed particle move in the direction
opposite to the field can always occur with finite probability.  This
would imply that the irreducibility and ergodicity proofs given in
Ref.~\cite{ToBiFi} may be extended to the NESS, provided that the
field intensity $E$ is finite.  Some technical steps, however, may be
not straightforward because, in contrast with the equilibrium case and
with the ASEP, the probability weight of constant-density
configurations is nonuniform in the NESS, even though the average
density profile is flat.

A first insight in the flow behavior can be gained by looking at the
particle current $J$ vs the driving force.  At densities below $\rho
\simeq 0.79$, $J$ first grows linearly with $E$ (ohmic regime) and
then tends to saturate at large fields.  This behavior is
qualitatively similar to that observed in the ASEP and will not be
further considered here.  At densities above $\rho \simeq 0.79$ the
transport properties differ rather markedly from the ASEP: the current
shows a transition from the ohmic regime at small fields to a
non-monotonic regime in which the current attains a maximum at finite
field (rather than at saturation as in the previous case), and then
decreases for larger field, see fig.~1.
The region with negative slope in the ``current-voltage''
characteristics is traditionally known as incremental negative
resistance (NR) and is a key ingredient in many biological systems and
solid-state devices.  Microscopic stochastic models yielding NR are
known~\cite{NR,Se_ratchet}.  In our case, NR does not depend on any
static interaction and occurs because, at high density and increasing
field, the particle rearrangements needed to remove obstruction to the
flow, require more and more particle moves against, or normal to, the
field direction (see, fig.~4c in Ref.~\cite{Juanpe} for an
illustration).
Three distinct behaviors may arise at high~$\rho$:

\begin{itemize}
\item[I.] Saturation current attains a finite (nonzero) value.

\item[II.] Saturation current vanishes.  

\item[III.] Current vanishes at a finite driving force.
\end{itemize}
Numerical results suggest that regime I occurs in the range $0.79
<\rho < 0.83$ while regime II appears at higher density.  For the
present model, the ergodicity result mentioned above should prevent
the existence of regime III in the thermodynamic limit, for large but
finite $E$.  Distinguishing between the two latter regimes, however,
may be difficult due to the strong finite-size effects related to
bootstrap percolation~\cite{ToBiFi,Paolo}.  The characterization of
these effects is notoriously difficult and will not be attempted here.
Kinetic constraints having a more involved dependence on the
neighboring particles~\cite{BiTo,JeSc}, will arguably have a jamming
transition at finite field in the thermodynamic limit.  Moreover, KCMs
having single mobility defects may only have type-I behavior, no
matter how large the particle density. This is confirmed by the
behavior of a noncooperative driven KA model on triangular
lattice.

\begin{figure}
 \begin{center}
  \includegraphics[width=6.5cm]{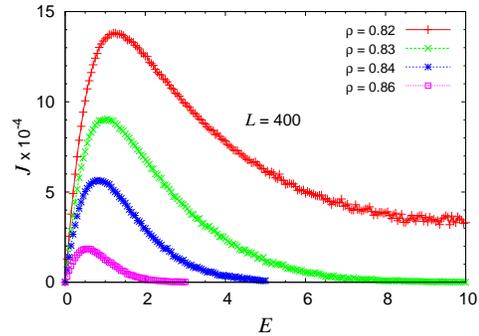}
\end{center}
\caption{Stationary current $J$ vs field $E$ at density $\rho$.}
\label{Fig:J_L400}
\end{figure}

NESS dynamics is expected to be slow in the NR region of high density
as the current becomes vanishingly small at increasing forcing.  This
can be analyzed through the persistence function $\phi(t)$, i.e. the
probability that the occupation variable of a lattice site has never
changed between times 0 and $t$.  The two stages responsible for
equilibrium relaxation -- which consist of particle rearrangements
within spatially extended mobility defects and their subsequent
coalescence~\cite{ToBiFi} -- are found to respond differently to the
applied field.

\begin{figure}
\begin{center}
  \includegraphics[width=6.5cm]{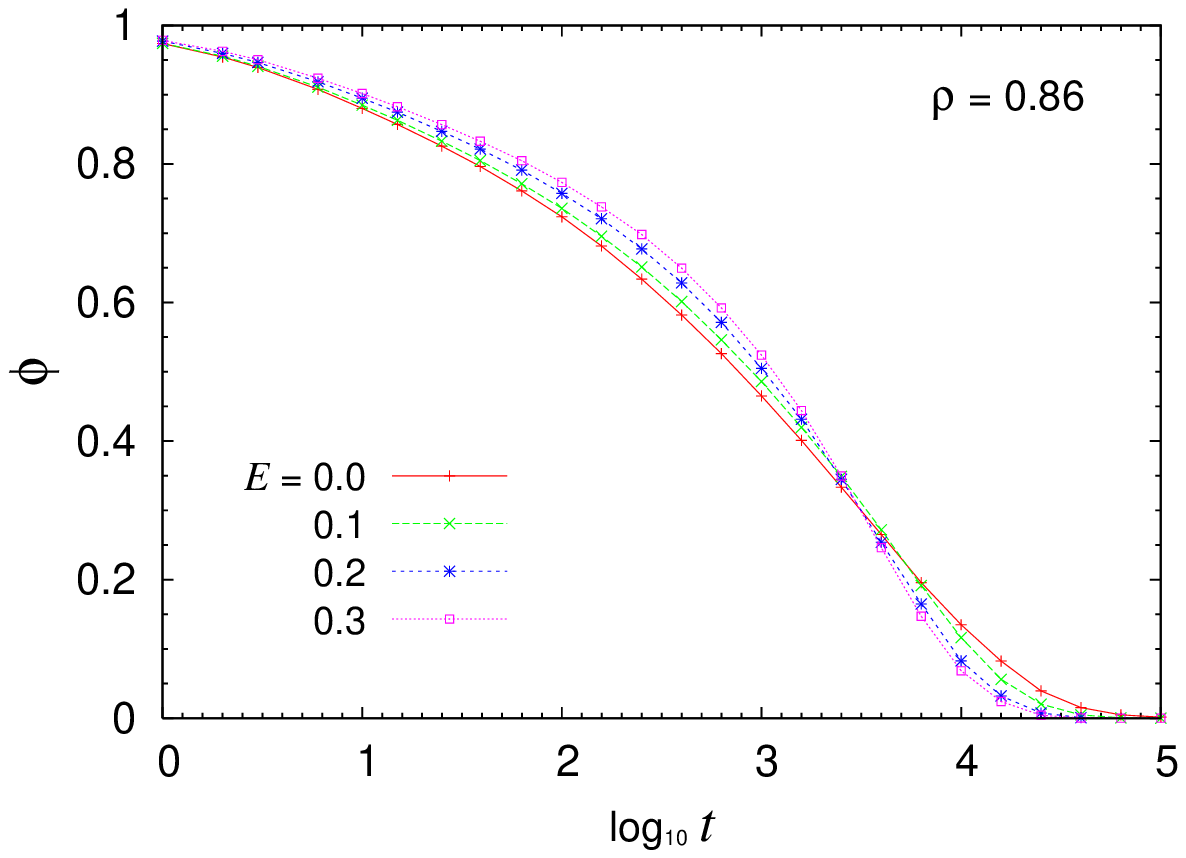}
  \includegraphics[width=6.5cm]{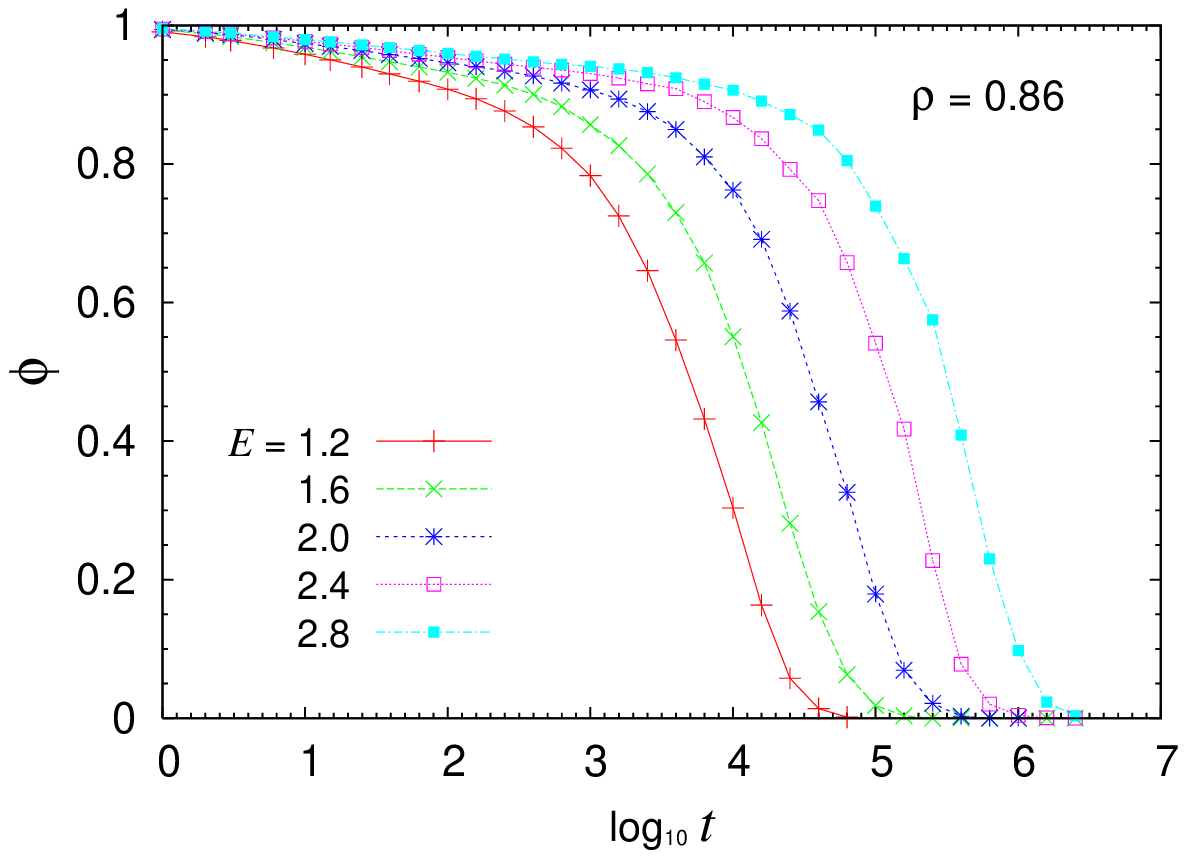}
\end{center}
\caption{Steady state persistence  $\phi$ vs time $t$ at 
  density $\rho$, for small (upper panel) and large (lower panel) 
  applied field $E$. 
}
\label{Fig:p_rho0.86}

\end{figure}

At small field, relaxation under flow is slower than at equilibrium on
short and intermediate time scales.  At longer time scales, on the
other hand, the macro-defects coalescence speeds up, and the
relaxation behavior changes smoothly from a stretched to a simple
exponential decay.  The two stages of relaxation are well separated by
a density-dependent crossover time, see fig.~\ref{Fig:p_rho0.86}
(upper panel).

\begin{figure}
\begin{center}
  \includegraphics[width=6.5cm]{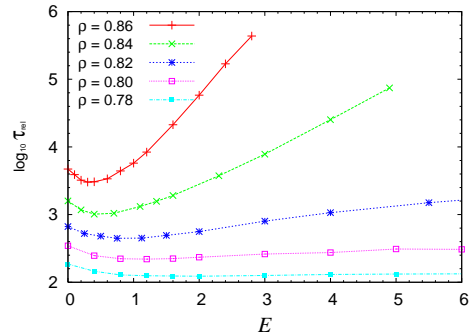}
\end{center}
\caption{Integrated relaxation time $\taurel$ vs driving force $E$.}
\label{Fig:tau}
\end{figure}

At large field, the thinning regime disappears because the time over
which the first-stage relaxation occurs in the NESS exceeds the one
required for the second-stage to take place in equilibrium.
Consequently, $\phi$ develops a more marked two-step behavior in which
the correlation decay becomes slower and slower at increasing field.
The resulting thickening-like behavior is shown in the lower panel of
fig.~\ref{Fig:p_rho0.86}. The late stage relaxation in this regime
obeys the superposition principle, $\phi(t) =
\Phi(t/\taurel(\rho,E))$, where the scaling function $\Phi$ is a
simple exponential for the model under consideration, and
$\taurel(\rho,E) = \int_0^{\infty} \phi(t) dt$ is the integrated
relaxation time.
%
%
$\taurel$ is a measure of the system viscosity and encodes the overall
relaxation resulting from the superposition of the two above-mentioned
competing effects. Fig.~\ref{Fig:tau} clearly shows that at large
enough density, $\taurel$ first decreases at small field and then
increases (exponentially) at large field.
To better appreciate the result above one should consider that outside
the linear response regime the ``resistivity'' $E/J$ increases with
$E$, while the relaxation time is roughly independent of $E$ at low
density.  This confirms that viscosity (rather than ``resistivity'')
does generally account for the genuine flow properties of our model.
The nonmonotonic dependence of viscosity at high density and the
related transition from thinning to thickening behavior, are features
observed in sheared hard-sphere suspensions~\cite{Larson}.  In fact,
flow curve analogs, obtained by identifying the shear stress with $E$
and the shear rate with $E/\taurel(\rho,E)$, are similar to those
found in the ``Model II'' of the schematic mode-coupling theories
studied in~\cite{Holmes}. In this case, since kinetic constraints are
``hard'', no fluidisation occurs after jamming no matter how strong
the force may be. If violation of constraints is allowed at large
field, however, one should recover the flow curves of ``Model I'' and
``Model III'' of Ref.~\cite{Holmes}.

Notice that in contrast with the nonmonotonic behavior of viscosity
and current, the fraction of blocked particles increases with $E$,
even when transport is ohmic.  The driving force thus generally
enhances the clustering of particles and appears to be akin to a
static short-range attraction. In fact, models of this type do have a
nonmonotonic dependence of the relaxation time with the attraction
strength at equilibrium~\cite{Geissler}. This observation suggests an
analogy between attraction-driven inverse freezing and
shear-thickening/jamming.

\begin{figure}
\begin{center}
  \includegraphics[width=6.5cm]{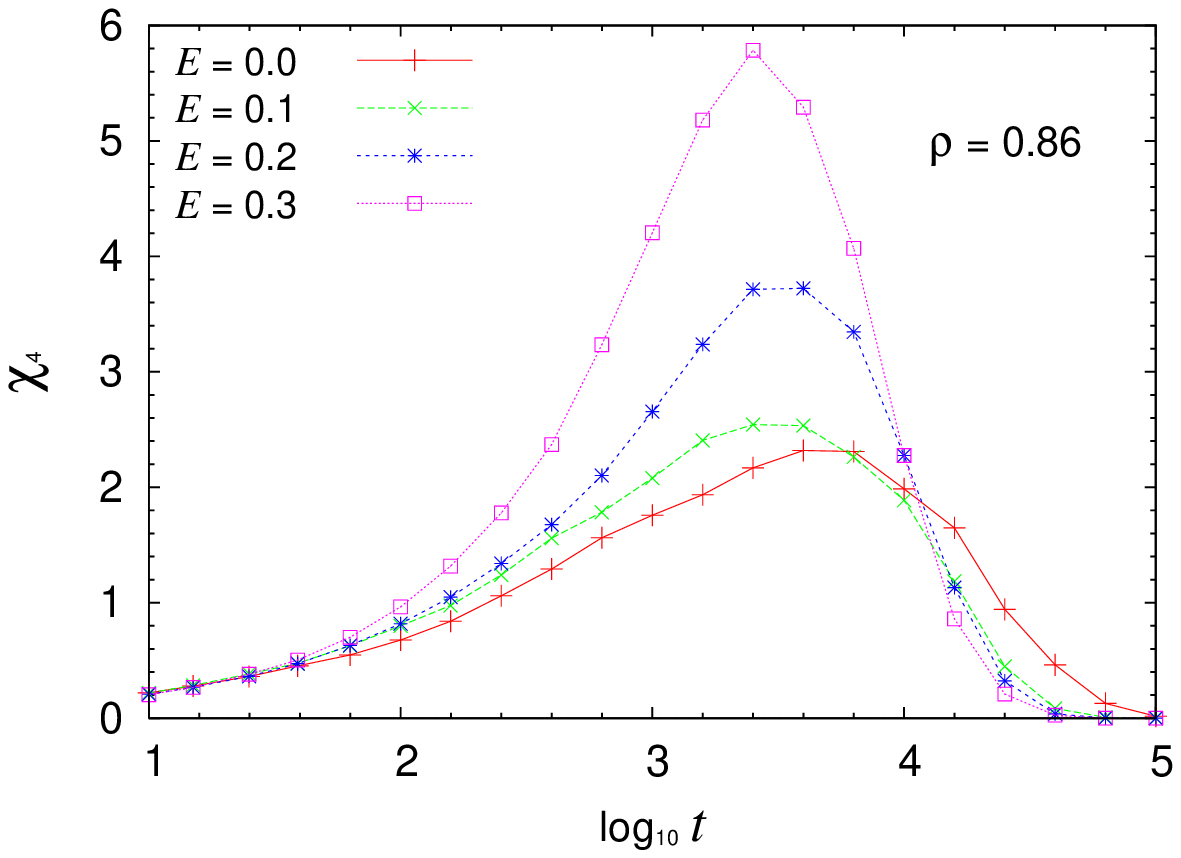}
  \includegraphics[width=6.5cm]{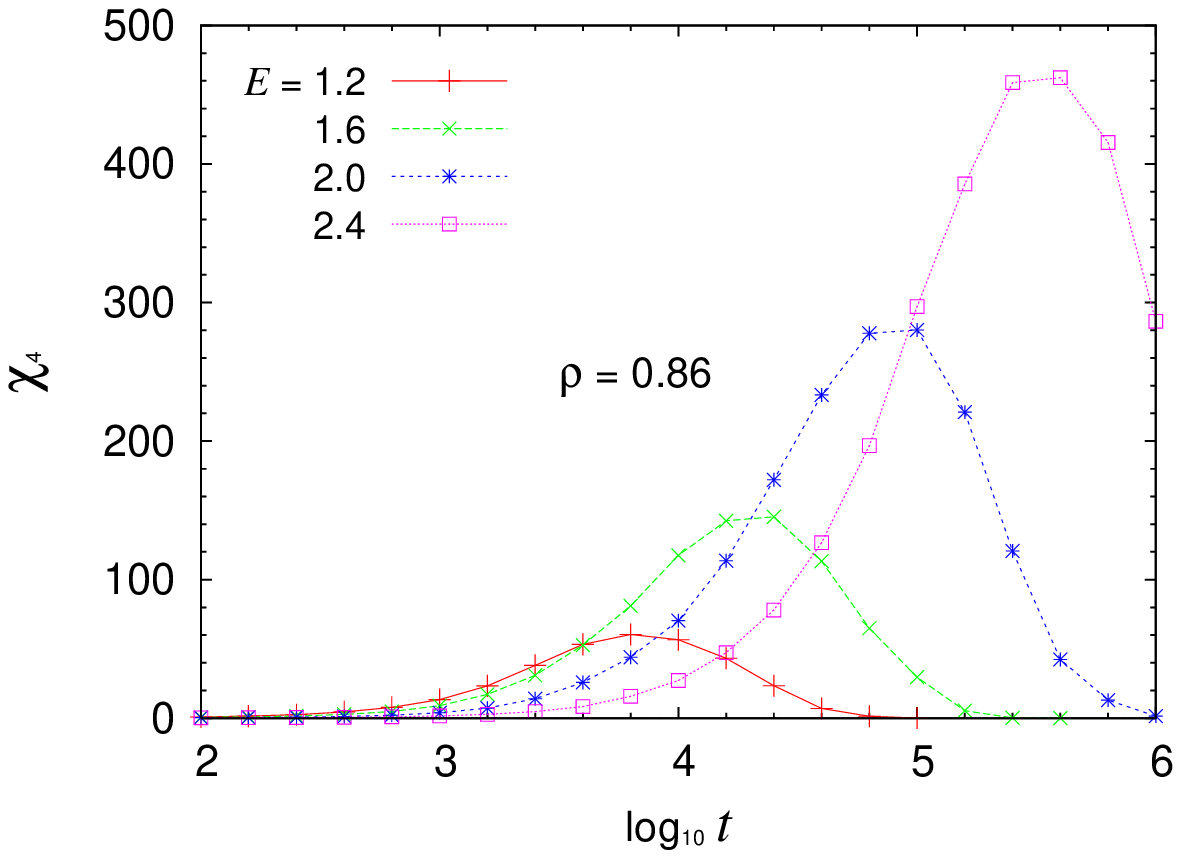}
\end{center}
\caption{Steady state four-point correlation $\chi_4$ vs time $t$ for
  small (upper panel) and large (lower panel) 
  applied field $E$. 
}
\label{Fig:chi4_rho0.86}
\end{figure}

Finally, dynamic heterogeneity in the NESS can be quantified by the
fluctuations of persistence, $\chi_4(t) = N(\langle \phi^2(t)\rangle -
\langle \phi(t)\rangle^2)$ where $N=\rho V$~\cite{chi4,Ludo_science}.
Similarly to equilibrium supercooled liquids~\cite{Ludo_science} we
find that dynamic heterogeneity and long range order grows with the
applied field, as exemplified by the increasing peak in the dynamic
susceptibility $\chi_4(t)$, and consistently with the monotonic
increase of the fraction of blocked particle with $E$.  However, in
the thinning regime the peak occurs at shorter and shorter times as
the field grows.  In the thickening regime, on the other hand, the
peak height and the time at which the peak occurs both increases with
the applied field, as shown in fig.~\ref{Fig:chi4_rho0.86}.
Interestingly, even though the dynamics is obviously nonisotropic, the
longitudinal and transverse persistence (and related fluctuations) are
almost indistinguishable.  Tiny differences only appear in the early
stage of relaxation at large field.  Thus, there is no substantial
change in the corresponding longitudinal and transverse relaxation
times.

\paragraph{Conclusions}

To summarize, when KCMs are driven into a nonequilibrium stationary
state they show both an initial speed-up of the dynamics and then a
pronounced slowdown at increasing fields. This suggests a common
kinetic mechanism for the nonmonotonic viscosity of athermal systems
and their thinning-thickening transition.
The basic ingredient is the presence of two relaxation stages
responding differently to the applied field.  Such a behavior occurs
in the simplest case of a spatially uniform nonconservative force.  We
expect that including a space dependent driving (to mimick more
realistically a shear stress) yields similar results.  In particular,
the peculiar time evolution of dynamic heterogeneities observed here
in the thinning and thickening regime should be experimentally
accessible in sheared granular materials and concentrated colloidal
suspensions.  Finally, the existence of a jamming/blocking transition,
possibly induced by strong finite-size effects, implies that driven
cooperative KCMs, are able to sustain indefinitely an applied stress
above a certain threshold, and thus to provide a microscopic
realization of {\em fragile matter}~\cite{fragilematter}.


\begin{thebibliography}{35}
  
\bibitem{ToBiFi} C. Toninelli, G. Biroli and D. Fisher, J. Stat. Phys.
  {\bf 120} 167 (2005)

\bibitem{Paolo} P. De Gregorio, A. Lawlor, P. Bradley, and K. Dawson,
  Proc. Natl.  Acad. Sci. USA {\bf 102}, 5669 (2005)

\bibitem{Merolle} M. Merolle, J.P. Garrahan, D. Chandler, Proc. Natl.
  Acad. Sci. USA {\bf 102}, 10837 (2005)

\bibitem{KCM_reviews} F. Ritort and P. Sollich, Adv. Phys. {\bf 52},
  219 (2003).

\bibitem{Se_review} M. Sellitto, J. Phys.: Condens. Matter {\bf 14},
  1455 (2002).

\bibitem{SeBiTo} M. Sellitto, G. Biroli and C. Toninelli, Europhys.
  Lett. {\bf 69}, 496 (2005).

\bibitem{Larson} R. G. Larson, {\sl The Structure and Rheology of
    Complex Fluids}, (OUP, Oxford, 1999), Chap.~6, and refs. therein.


\bibitem{BeBaKu} L. Berthier, J.-L. Barrat and J. Kurchan, Phys. Rev.
  E {\bf 61}, 5464 (2000);  K. Miyazaki, D.R. Reichman, Phys. Rev. E
  {\bf 66}, 050501 (2002);  M. Fuchs, M.E. Cates, Phys. Rev. Lett.
  {\bf 89}, 248304 (2002)

\bibitem{bibette} E. Bertrand, J. Bibette and V. Schmitt, Phys. Rev. E
  {\bf 66}, 060401(R) (2002)

\bibitem{note_jamming} The term jamming denotes here a dynamic arrest
  transition induced by nonconservative forces (rather than by
  temperature or pressure, as in the usual glass transition).

\bibitem{Holmes} See: C. B. Holmes, M. Fuchs, M. E. Cates and P.
  Sollich, J. Rheol. {\bf 49}, 237 (2005)

\bibitem{SeKu} M. Sellitto and J. Kurchan, Phys. Rev. Lett. {\bf 95},
  236001 (2005); M. Sellitto, Phys. Rev. B {\bf 73}, 180202(R) (2006)

\bibitem{Bonn} A. Fall, N. Huang, F. Bertrand, G. Ovarlez and D. Bonn,
  Phys. Rev. Lett. {\bf 100}, 018301 (2008)

\bibitem{Roberta} R. Angelini, G. Ruocco, S. De Panfilis and F. Sette,
  arXiV: 0802.4046

\bibitem{ASEP_reviews} For a review, see G.M. Sch\"utz, in {\em Phase
    Transitions and Critical Phenomena}, Vol. 19, C. Domb and J.
  Lebowitz eds. (Academic Press, London, 2000).

\bibitem{KoAn} W. Kob and H.C. Andersen, Phys. Rev. E {\bf 48}, 4364
  (1993). J. J\"ackle and A. Kr\"onig, J. Phys.: Condens. Matter  {\bf 6},
  7633 (1994)

\bibitem{acid} J.T. MacDonald, J.H. Gibbs and A. Pipkin, Biopolymers
  {\bf 6}, 1 (1968)

\bibitem{Derrida_review} B. Derrida, J. Stat. Mech.  P07023 (2007)

\bibitem{Andreas} For a review, see D. Chowdhury, A. Schadschneider,
  and K. Nishinari, Phys. of Life Rev. {\bf 2}, 318 (2005)

\bibitem{Se_ratchet} M. Sellitto, Phys. Rev. E {\bf 65}, 020101 (2002)

\bibitem{NR} V. Balakrishnan and C. Van den Broeck, Physica A {\bf
    217}, (1995); C. Maes and W. Vanderpoorten, Phys.  Rev. B {\bf
    53}, 12889 (1996); G.A. Cecchi and M.O. Magnasco, Phys.  Rev.
  Lett. {\bf 76}, 1968 (1996); B. Cleuren and C. Van den Broeck, Phys.
  Rev. E {\bf 65}, 030101(R) (2002)

\bibitem{Juanpe} R.L. Jack, D. Kelsey, J.P. Garrahan, and D. Chandler,
  arXiv:0803.2002

\bibitem{BiTo} G. Biroli and C. Toninelli, arXiv:0709.0583

\bibitem{JeSc} M. Jeng and J.M. Schwartz, arXiv:0708.0582

\bibitem{Geissler} P.L. Geissler, and D.R. Reichman, Phys. Rev. E
  {\bf 71}, 031206 (2005)

\bibitem{chi4} S. Franz and G. Parisi, J. Phys.: Condens. Matter {\bf 12}
  6335 (2000)
\bibitem{Ludo_science} L. Berthier, G. Biroli, J.-P.  Bouchaud, L.
  Cipelletti, D. El Masri, D. L'H\^ote, F. Ladieu, M. Pierno, Science
  {\bf 310}, 1797 (2005); L. Berthier, G. Biroli, J.-P. Bouchaud, W.
  Kob, K.  Miyazaki, D.R. Reichman, J. Chem. Phys. {\bf 126}, 184504
  (2007)

\bibitem{fragilematter} M.E. Cates, J.P. Wittmer, J.-P.~Bouchaud and
  P.~Claudin, Phys. Rev. Lett. {\bf 81}, 1841 (1998)

\end{thebibliography}
\end{document}